\documentclass[english]{sbrt}
\usepackage{xcolor}
\usepackage{cite}
\usepackage[english]{babel}
\usepackage[utf8]{inputenc}
\usepackage{graphicx}           
\usepackage{amsmath}   
\usepackage{amssymb}      
\usepackage{microtype}   
\usepackage{tikz}
\usepackage{bm}
\usepackage{booktabs}
\usepackage{capt-of}
\usepackage[ruled,linesnumbered]{algorithm2e}
\SetKwProg{Fn}{begin}{}{end}
\SetKw{And}{and}
\SetAlgoNlRelativeSize{-1}
\SetNlSty{textbf}{}{:}

\begin{document}

\title{Structured Channel Parameter Estimation for RIS-Assisted MIMO Systems via Kronecker and Tensor Factorizations}

\author{Melryllin G. O. Sousa, Jean C. S. Ferreira and Fazal-E Asim
\thanks{\fontsize{6.3}{7.938}\selectfont Melryllin G. O. Sousa, Department of Teleinformatics Engineering, Federal University of Ceará (UFC), Fortaleza-CE, e-mail: mryllin@alu.ufc.br; Jean C. S. Ferreira, Department of Teleinformatics Engineering, Federal University of Ceará (UFC), Fortaleza-CE, e-mail: jeancarlos8249@alu.ufc.br; Fazal Asim, Department of Teleinformatics Engineering, Federal University of Ceará (UFC), Fortaleza-CE, e-mail: fazalasim@ufc.br. This work is partially supported by the National Institute of Science and Technology (INCT-Signals) sponsored by Brazil’s National Council for Scientific and Technological Development (CNPq) (Proc. 406517/2022-3), and FUNCAP (Proc. INCT-25255-82587.32.41/64).}
}

\maketitle

\markboth{XLIV BRAZILIAN SYMPOSIUM ON TELECOMMUNICATIONS AND SIGNAL PROCESSING - SBrT 2026, SEPTEMBER 29TH TO OCTOBER 2ND, 2026, SALVADOR, BA}{}

\begin{abstract}
This paper proposes structured channel-parameter estimators for reconfigurable intelligent surface (RIS)-assisted multiple-input multiple-output (MIMO) systems employing uniform rectangular arrays under line-of-sight propagation. By exploiting the Kronecker and multilinear structures of the cascaded channel, we introduce the Successive Kronecker Factorization (SKF) alongside two Third-Order Tensor Parameter Estimation (TOPE) methods: TOPE-ALS and TOPE-HOSVD. These estimators recover spatial signatures from the pilot-filtered signal matrix, and then decouple the corresponding horizontal and vertical spatial frequencies. This structured formulation achieves benchmark estimation accuracy while reducing computational complexity, offering a efficient solution for RIS-aided systems.
\end{abstract}
\section{Introduction}
Reconfigurable intelligent surfaces (RISs) enhance wireless coverage and spectral efficiency through programmable reflecting elements~\cite{ref1,ref2}. However, these gains in multiple-input multiple-output (MIMO) architectures rely on accurate channel state information. This estimation is challenging because the cascaded base station (BS)--RIS--user equipment (UE) channel and the massive number of passive RIS elements incur a prohibitive training overhead for conventional unstructured estimators~\cite{ref3,ref4}.

To reduce this burden, recent works exploit structural priors in millimeter-wave line-of-sight (LoS) scenarios~\cite{ref5,ref6}.  By leveraging the Kronecker structure of uniform rectangular arrays, geometry-aware methods bypass unstructured high-dimensional estimation, drastically reducing the number of unknowns under limited pilot resources. Furthermore, tensor models inherently preserve the multidimensional coupling among the BS, RIS, and UE, facilitating accurate spatial-signature extraction via algebraic decompositions~\cite{ref7,ref8,ref9,ref10,ref11,ref15}.

However, state-of-the-art estimators relying on these priors often face a strict accuracy-complexity tradeoff. For instance, recent approaches like hybrid Kronecker factorization and multi-rank-one approximations (HKMR) and two-stage higher-dimensional rank-one approximations (TSHDR)~\cite{ref12} either sacrifice accuracy for low computational complexity or achieve near-optimal performance at a prohibitive algorithmic cost. To overcome this, this paper proposes three accurate and efficient channel estimation methods for RIS-assisted MIMO systems employing URAs under a single-path dominant LoS scenario. The main contributions are:

\begin{itemize}
    \item We propose three estimators: Successive Kronecker Factorization (SKF), Third-Order Tensor Parameter Estimation (TOPE) with alternating least squares (TOPE-ALS) and with higher-order singular value decomposition (TOPE-HOSVD). By strategically altering the factorization sequence to isolate the BS, RIS, and UE spatial signatures prior to the horizontal/vertical frequency decoupling, these methods effectively break the accuracy-complexity trade-off.
    \item Supported by the complexity analysis, we demonstrate that the proposed estimators achieve the near-optimal accuracy of the TSHDR~\cite{ref12} benchmark while maintaining a minimal computational footprint comparable to the suboptimal HKMR~\cite{ref12} baseline.
    \item Unlike prior works [12] that derived the Cramér-Rao Lower Bound (CRLB) based solely on the intermediate pilot-filtered signal, yielding a loose bound that does not fully capture the system's true statistical limits, we derive the exact CRLB. By explicitly incorporating the transmit pilot and RIS phase-shift matrices into the Fisher Information Matrix, we establish the true theoretical error bound for the complete cascaded LoS channel. This rigorous benchmark demonstrates that our proposed estimators match the near-optimal accuracy of the TSHDR~\cite{ref12} algorithm across almost all SNR regimes, operating closely to the fundamental theoretical limits.
    
\end{itemize}

The remainder of the paper is organized as follows. Section~II introduces the geometric and signal models. Section~III presents the proposed SKF, TOPE-ALS, and TOPE-HOSVD estimators together with their complexity analysis. Section~IV presents the Cramér-Rao Lower Bound. Section~V reports the numerical results, and Section~VI concludes the paper.

\textbf{\textit{Notation:}} Vectors, matrices, and tensors are denoted by $\bm{a}$, $\bm{A}$, and $\mathcal{X}$. Transpose, Hermitian, and inverse operators are $(\cdot)^T$, $(\cdot)^H$, and $(\cdot)^{-1}$. Kronecker, Hadamard, Khatri-Rao, and outer products are $\otimes$, $\odot$, $\diamond$, and $\circ$. The operators $\operatorname{vec}(\cdot)$ and $\operatorname{diag}(\cdot)$ denote matrix vectorization and diagonalization. The $n$-mode product of $\mathcal{X} \in \mathbb{C}^{I_1 \times I_2 \times I_3}$ with matrix $\bm{A}$ is $\mathcal{X} \times_n \bm{A}$. Complex numbers are $\mathbb{C}$.

\section{System Model}
\begin{figure}[htbp]
    \centering
    \includegraphics[width=0.35\textwidth]{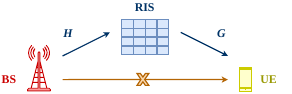}
    \caption{\label{fig:figura}Illustration of the considered RIS-assisted MIMO communication system, where both the BS and UE employ URAs.}
\end{figure}
Consider the RIS-assisted MIMO communication system illustrated in Fig.~\ref{fig:figura}, where a base station (BS) equipped with $M$ antennas communicates with a user equipment (UE) equipped with $Q$ antennas through an $N$-element reconfigurable intelligent surface (RIS). The BS, RIS, and UE are assumed to employ uniform rectangular arrays (URAs), with $M=M_yM_z$, $N=N_yN_z$, and $Q=Q_yQ_z$ antenna or reflecting elements. We focus on a millimeter-wave propagation scenario in which the BS--RIS and RIS--UE links are dominated by a single line-of-sight (LoS) component, while the direct BS--UE link is assumed to be blocked.

Let $\boldsymbol{H} \in \mathbb{C}^{N \times M}$ denote the BS--RIS channel and $\boldsymbol{G} \in \mathbb{C}^{Q \times N}$ denote the RIS--UE channel. Under the single-path geometric model, these channels are written as
\begin{equation}
    \boldsymbol{H} = \alpha \, \boldsymbol{b}(\phi_{\text{ris}_A}, \theta_{\text{ris}_A}) \, \boldsymbol{a}^{\text{T}}(\phi_{\text{bs}}, \theta_{\text{bs}}) \in \mathbb{C}^{N \times M},
\end{equation}
\begin{equation}
    \boldsymbol{G} = \beta \, \boldsymbol{q}(\phi_{\text{ue}}, \theta_{\text{ue}}) \, \boldsymbol{p}^{\text{T}}(\phi_{\text{ris}_D}, \theta_{\text{ris}_D}) \in \mathbb{C}^{Q \times N},
\end{equation}
where $\alpha,\beta \in \mathbb{C}$ are the complex channel gains. The vectors $\boldsymbol{a}(\cdot) \in \mathbb{C}^{M \times 1}$ and $\boldsymbol{q}(\cdot) \in \mathbb{C}^{Q \times 1}$ denote the BS transmit and UE receive steering vectors, respectively. Similarly, $\boldsymbol{b}(\cdot) \in \mathbb{C}^{N \times 1}$ and $\boldsymbol{p}(\cdot) \in \mathbb{C}^{N \times 1}$ denote the RIS steering vectors associated with the arrival from the BS and the departure toward the UE, respectively. The angles $(\phi,\theta)$ represent the azimuth and elevation directions associated with each array response.
\subsection{Channel Factorization}
The URA geometry induces a separable steering-vector structure, which is exploited by the proposed estimators. For example, assume that the BS array lies in the $y$-$z$ plane with half-wavelength inter-element spacing. Let $m_y \in \{1,\ldots,M_y\}$ and $m_z \in \{1,\ldots,M_z\}$ denote the horizontal and vertical indices of the $m$-th BS antenna, respectively. The corresponding entry of the BS steering vector is given by $[\boldsymbol{a}(\phi_{\text{bs}}, \theta_{\text{bs}})]_m = e^{-j \left[ (m_y - 1) \mu_{\text{bs}} + (m_z - 1) \psi_{\text{bs}} \right]}$, where $\mu_{\text{bs}} = \pi \sin \theta_{\text{bs}} \sin \phi_{\text{bs}}$ and $\psi_{\text{bs}} = \pi \cos \theta_{\text{bs}}$ are the horizontal and vertical spatial frequencies, respectively (see \cite{ref12} for further details on this structure). Hence, the BS steering vector can be decomposed as $ \boldsymbol{a}(\phi_{\text{bs}}, \theta_{\text{bs}}) = \boldsymbol{a}_y(\mu_{\text{bs}}) \otimes \boldsymbol{a}_z(\psi_{\text{bs}})$, where $\boldsymbol{a}_y(\mu_{\text{bs}}) \in \mathbb{C}^{M_y \times 1}$ and $\boldsymbol{a}_z(\psi_{\text{bs}}) \in \mathbb{C}^{M_z \times 1}$ are one-dimensional steering vectors along the $y$ and $z$ axes, respectively.

The same decomposition applies to the RIS arrival, RIS departure, and UE steering vectors. Therefore, by using the mixed-product property of the Kronecker product, the two channel matrices can be expressed as
\begin{equation}
    \boldsymbol{H} = \alpha \left[ \boldsymbol{b}_y(\mu_{\text{ris}_A}) \boldsymbol{a}_y^{\text{T}}(\mu_{\text{bs}}) \right] \otimes \left[ \boldsymbol{b}_z(\psi_{\text{ris}_A}) \boldsymbol{a}_z^{\text{T}}(\psi_{\text{bs}}) \right],
\end{equation}
\begin{equation}
    \boldsymbol{G} = \beta \left[ \boldsymbol{q}_y(\mu_{\text{ue}}) \boldsymbol{p}_y^{\text{T}}(\mu_{\text{ris}_D}) \right] \otimes \left[ \boldsymbol{q}_z(\psi_{\text{ue}}) \boldsymbol{p}_z^{\text{T}}(\psi_{\text{ris}_D}) \right].
\end{equation}

\subsection{Signal Model}

Let $\boldsymbol{S} \in \mathbb{C}^{M \times T}$ denote the BS pilot matrix, where $T$ is the pilot length. In this work, $\boldsymbol{S}$ is chosen with a Hadamard structure. During the $k$-th training block, the RIS applies the reflection vector $\boldsymbol{\omega}_k \in \mathbb{C}^{N \times 1}$. The received pilot signal at the UE is then modeled as
\begin{equation}
    \boldsymbol{X}_k = \boldsymbol{G} \operatorname{diag}(\boldsymbol{\omega}_k) \boldsymbol{H} \boldsymbol{S} + \boldsymbol{V}_k \in \mathbb{C}^{Q \times T},
\end{equation}
where $\boldsymbol{V}_k \in \mathbb{C}^{Q \times T}$ is the additive white Gaussian noise matrix with independent entries distributed as $\mathcal{CN}(0,\sigma_n^2)$. The RIS training matrix is denoted by $\boldsymbol{\Omega}=[\boldsymbol{\omega}_1,\ldots,\boldsymbol{\omega}_K] \in \mathbb{C}^{N \times K}$ and is selected from a normalized DFT codebook such that $\boldsymbol{\Omega}\boldsymbol{\Omega}^H=\boldsymbol{I}_N$. Defining the effective cascaded MIMO channel during the $k$-th training block as $\boldsymbol{U}_k = \boldsymbol{G} \operatorname{diag}(\boldsymbol{\omega}_k) \boldsymbol{H} \in \mathbb{C}^{Q \times M}$, the received signal can be compactly rewritten as
\begin{equation}
    \boldsymbol{X}_k = \boldsymbol{U}_k \boldsymbol{S} + \boldsymbol{V}_k.
    \label{eq:rec_signal}
\end{equation}

   \section{Channel Parameter Estimation Methods}
This section presents three structured methods for estimating the channel parameters from the pilot-filtered observations. The first method, referred to as Successive Kronecker Factorization (SKF), directly exploits the hierarchical Kronecker structure induced by the URA geometry. The other two methods first reshape the pilot-filtered matrix into a third-order tensor and then extract the BS, RIS, and UE spatial signatures using either Alternating Least Squares (TOPE-ALS) or Higher-Order Singular Value Decomposition (TOPE-HOSVD). After this first stage, all methods use the same least-squares Kronecker factorization (LSKronF) principle to separate the horizontal and vertical spatial-frequency components.

\subsection{Pilot-Filtered Signal Matrix}
We first derive the pilot-filtered signal matrix, which is the common input to all proposed estimators. Assuming that the pilot matrix satisfies $\boldsymbol{S}\boldsymbol{S}^H=\boldsymbol{I}_M$, right-multiplying \eqref{eq:rec_signal} by $\boldsymbol{S}^H$ gives
\begin{equation}
\widetilde{\boldsymbol{U}}_k = \boldsymbol{X}_k \boldsymbol{S}^{H} = \boldsymbol{G} \operatorname{diag}(\boldsymbol{\omega}_k) \boldsymbol{H} + \widetilde{\boldsymbol{V}}_k \in \mathbb{C}^{Q \times M},
\label{eq:pilot_filtered_block}
\end{equation}
where $\widetilde{\boldsymbol{V}}_k=\boldsymbol{V}_k\boldsymbol{S}^H$ is the filtered noise term. Vectorizing \eqref{eq:pilot_filtered_block} and collecting the $K$ training blocks into $\boldsymbol{U}=[\boldsymbol{u}_1,\ldots,\boldsymbol{u}_K]$, with $\boldsymbol{u}_k=\operatorname{vec}(\widetilde{\boldsymbol{U}}_k)$, and applying the property $\operatorname{vec}(\boldsymbol{A}\operatorname{diag}(\boldsymbol{b})\boldsymbol{C})=(\boldsymbol{C}^T\diamond\boldsymbol{A})\boldsymbol{b}$ yields
\begin{equation}
    \boldsymbol{U} = (\boldsymbol{H}^T \diamond \boldsymbol{G}) \boldsymbol{\Omega} + \widetilde{\boldsymbol{V}} \in \mathbb{C}^{QM \times K}.
    \label{eq:stacked_filtered_blocks}
\end{equation}
Applying matched filtering with the known RIS training matrix and defining $\boldsymbol{V}_{E}=\widetilde{\boldsymbol{V}}\boldsymbol{\Omega}^H$ gives
\begin{equation}
    \boldsymbol{E} = \boldsymbol{U} \boldsymbol{\Omega}^H = \boldsymbol{H}^T \diamond \boldsymbol{G} + \boldsymbol{V}_{E} \in \mathbb{C}^{QM \times N}.
    \label{eq:E_observation}
\end{equation}

Substituting the single-path channel models into \eqref{eq:E_observation}, the noiseless component admits the rank-one factorization
\begin{equation}
    \boldsymbol{H}^T \diamond \boldsymbol{G} = \gamma (\boldsymbol{a} \otimes \boldsymbol{q})\boldsymbol{n}^T,
    \label{eq:kr_channel_factorization}
\end{equation}
where $\gamma=\alpha\beta$ is the cascaded complex gain and $\boldsymbol{n} = \boldsymbol{b} \odot \boldsymbol{p} \in \mathbb{C}^{N \times 1}$ is the combined RIS spatial signature. By exploiting the Kronecker structure of the planar arrays, the spatial frequencies decouple directly as $\boldsymbol{n}(\mu_y, \psi_z) = \boldsymbol{n}_y(\mu_y) \otimes \boldsymbol{n}_z(\psi_z)$, with components $\boldsymbol{n}_y(\mu_y) = \boldsymbol{b}_y(\mu_{\text{ris}_A}) \odot \boldsymbol{p}_y(\mu_{\text{ris}_D})$ and $\boldsymbol{n}_z(\psi_z) = \boldsymbol{b}_z(\psi_{\text{ris}_A}) \odot \boldsymbol{p}_z(\psi_{\text{ris}_D})$. Since the Hadamard product of two steering vectors adds their spatial frequencies, we have $\mu_y = \mu_{\text{ris}_A} + \mu_{\text{ris}_D}$ and $\psi_z = \psi_{\text{ris}_A} + \psi_{\text{ris}_D}$. Defining the joint BS-UE spatial signature as $\boldsymbol{\ell}=\boldsymbol{a}\otimes\boldsymbol{q}\in\mathbb{C}^{QM\times 1}$, the observation matrix can be compactly written as

\begin{equation}
    \boldsymbol{\hat{E}} \approx \gamma\boldsymbol{\ell}\boldsymbol{n}^T
    \in \mathbb{C}^{QM\times N}.
    \label{eq:E_final}
\end{equation}
\subsection{Successive Kronecker Factorization (SKF)}
The SKF algorithm sequentially factorizes the observation matrix to isolate the 1D steering vectors. Exploiting the rank-one structure in \eqref{eq:E_final}, the first stage extracts the joint BS--UE signature and the combined RIS signature by solving

\begin{equation}
    \{\hat{\boldsymbol{\ell}}, \hat{\boldsymbol{n}}\} = \arg \min_{\boldsymbol{\ell}, \boldsymbol{n}} \| \boldsymbol{E} - \boldsymbol{\ell} \boldsymbol{n}^T \|_F^2.
    \label{eq:skf_1}
\end{equation}
This is solved via the dominant rank-one approximation of $\boldsymbol{E}$, with the scalar gain and arbitrary scaling absorbed into the estimated factors.

Since $\boldsymbol{\ell}=\boldsymbol{a}\otimes\boldsymbol{q}$, the second stage reshapes or factorizes $\hat{\boldsymbol{\ell}}$ to separate the BS and UE two-dimensional steering vectors:
\begin{equation}
    \{\hat{\boldsymbol{a}}, \hat{\boldsymbol{q}}\} = \arg \min_{\boldsymbol{a}, \boldsymbol{q}} \| \hat{\boldsymbol{\ell}} - \boldsymbol{a} \otimes \boldsymbol{q} \|_2^2.
    \label{eq:skf_2}
\end{equation}

Finally, decomposing the 2D steering vectors into their horizontal and vertical 1D components yields three independent LSKronF problems:
\begin{equation}
    \{\hat{\boldsymbol{q}}_y, \hat{\boldsymbol{q}}_z\} = \arg \min_{\boldsymbol{q}_y, \boldsymbol{q}_z} \| \hat{\boldsymbol{q}} - \boldsymbol{q}_y \otimes \boldsymbol{q}_z \|_2^2,
    \label{eq:skf_3}
\end{equation}
\begin{equation}
    \{\hat{\boldsymbol{a}}_y, \hat{\boldsymbol{a}}_z\} = \arg \min_{\boldsymbol{a}_y, \boldsymbol{a}_z} \| \hat{\boldsymbol{a}} - \boldsymbol{a}_y \otimes \boldsymbol{a}_z \|_2^2,
    \label{eq:skf_4}
\end{equation}
\begin{equation}
    \{\hat{\boldsymbol{n}}_y, \hat{\boldsymbol{n}}_z\} = \arg \min_{\boldsymbol{n}_y, \boldsymbol{n}_z} \| \hat{\boldsymbol{n}} - \boldsymbol{n}_y \otimes \boldsymbol{n}_z \|_2^2.
    \label{eq:skf_5}
\end{equation}

The scalar ambiguities in \eqref{eq:skf_3}--\eqref{eq:skf_5} are eliminated by assuming the first unit entry of the steering vectors, leaving the joint complex path gain to be estimated in \eqref{eq:Complex_Path_Gain}. Subsequently, the final spatial frequencies are extracted from these normalized 1D steering vectors via Root-MUSIC or ESPRIT.

\subsection{Third-Order Tensor Parameter Estimation (TOPE)}
The TOPE methods first reshape the matched-filtered observation into a third-order tensor. The factors are then estimated either iteratively by ALS or as a closed form by HOSVD, followed by the same Kronecker-based horizontal/vertical decoupling and gain-estimation stages used by SKF.

To exploit the multilinear nature of the observation, TOPE-ALS and TOPE-HOSVD reshape the filtered matrix into a third-order tensor. In the noiseless case, vectorizing \eqref{eq:E_final} gives
\begin{equation}
    \bm{\hat{e}} \approx \gamma \bm{n}(\mu_y, \psi_z) \otimes \bm{a}(\mu_{\text{bs}}, \psi_{\text{bs}}) \otimes \bm{q}(\mu_{\text{ue}}, \psi_{\text{ue}}).
    \label{eq:evec_approx} \in \mathbb{C}^{QMN\times1}
\end{equation}

Let $\mathcal{\hat{E}} \in \mathbb{C}^{Q \times M \times N}$ be obtained by reshaping $\boldsymbol{\hat{e}}=\operatorname{vec}(\boldsymbol{\hat{E}})$ according to $[\mathcal{E}]_{q,m,n} \doteq [\boldsymbol{e}]_{(n-1)MQ + (m-1)Q + q}$, where $q = 1,\dots,Q$, $m = 1,\dots,M$, and $n = 1,\dots,N$. The corresponding noiseless tensor is rank one and can be written as

\begin{equation}
    \mathcal{\hat{E}} \approx \gamma\,\boldsymbol{q}(\mu_{\text{ue}}, \psi_{\text{ue}}) \circ \boldsymbol{a}(\mu_{\text{bs}}, \psi_{\text{bs}}) \circ \boldsymbol{n}(\mu_y, \psi_z).
    \label{eq:tensor_rank_one_model}
\end{equation}

We denote the mode-1, mode-2, and mode-3 unfoldings of $\mathcal{\hat{E}}$ by $[\widehat{\mathcal{E}}]_{(1)}$, $[\widehat{\mathcal{E}}]_{(2)}$, and $[\widehat{\mathcal{E}}]_{(3)}$, respectively. Both tensor-based methods use the model in \eqref{eq:tensor_rank_one_model}; they differ only in how the factors $\{ \hat{\boldsymbol{q}}, \hat{\boldsymbol{a}}, \hat{\boldsymbol{n}} \}$ are extracted.

For TOPE-ALS, the factors are obtained iteratively through conditional least-squares updates:
\begin{align}
    \hat{\bm{q}}^{(i)} &= \arg\min_{\bm{q}}
    \left\| [\widehat{\mathcal{E}}]_{(1)} - \bm{q}
    \left(\hat{\bm{n}}^{(i-1)} \diamond \hat{\bm{a}}^{(i-1)}\right)^T \right\|_F^2,\\
    \hat{\bm{a}}^{(i)} &= \arg\min_{\bm{a}}
    \left\| [\widehat{\mathcal{E}}]_{(2)} - \bm{a}
    \left(\hat{\bm{n}}^{(i-1)} \diamond \hat{\bm{q}}^{(i)}\right)^T \right\|_F^2,\\
    \hat{\bm{n}}^{(i)} &= \arg\min_{\bm{n}}
    \left\| [\widehat{\mathcal{E}}]_{(3)} - \bm{n}
    \left(\hat{\bm{a}}^{(i)} \diamond \hat{\bm{q}}^{(i)}\right)^T \right\|_F^2.
\end{align}

These updates proceed until the relative change of the squared reconstruction cost $\Gamma^{(i)}$ falls below a tolerance threshold $\varepsilon$, i.e., $|\Gamma^{(i)} - \Gamma^{(i-1)}| / \Gamma^{(i-1)} \le \varepsilon$, where $\Gamma^{(i)} = \min_{\gamma}\| \boldsymbol{e} - \gamma\hat{\boldsymbol{n}}^{(i)} \otimes \hat{\boldsymbol{a}}^{(i)} \otimes \hat{\boldsymbol{q}}^{(i)} \|_2^2$.

For TOPE-HOSVD, the same rank-one tensor approximation is solved non-iteratively in a closed form as:
\begin{equation}
    \{ \hat{\boldsymbol{q}}, \hat{\boldsymbol{a}}, \hat{\boldsymbol{n}} \} = \arg \min_{\boldsymbol{q}, \boldsymbol{a}, \boldsymbol{n},\gamma} \left\| \mathcal{E} - \gamma\boldsymbol{q} \circ \boldsymbol{a} \circ \boldsymbol{n} \right\|_{\text{F}}^2,
\end{equation}
where the factor estimates are obtained from the dominant left singular vectors of $[\widehat{\mathcal{E}}]_{(1)}$, $[\widehat{\mathcal{E}}]_{(2)}$, and $[\widehat{\mathcal{E}}]_{(3)}$. Thus, unlike the iterative ALS, HOSVD extracts these factors in a closed form fashion. Upon recovering $\hat{\boldsymbol{q}}$, $\hat{\boldsymbol{a}}$, and $\hat{\boldsymbol{n}}$, they are decoupled into 1D steering vectors via \eqref{eq:skf_3}--\eqref{eq:skf_5} to estimate the spatial frequencies using the same Root-MUSIC/ESPRIT backend.

\subsection{Complex Path Gain Estimation}

Once the horizontal and vertical spatial frequencies are successfully decoupled by any of the proposed algorithms (SKF, TOPE-ALS, or TOPE-HOSVD), the joint complex path gain can be calculated as
\begin{equation}
    \hat{\gamma} = \frac{1}{QMN} \mathcal{E} \times_1 \hat{\boldsymbol{q}}^H \times_2 \hat{\boldsymbol{a}}^H \times_3 \hat{\boldsymbol{n}}^H,
    \label{eq:Complex_Path_Gain}
\end{equation}
where $\hat{\boldsymbol{q}} = \hat{\boldsymbol{q}}_y \otimes \hat{\boldsymbol{q}}_z$, $\hat{\boldsymbol{a}} = \hat{\boldsymbol{a}}_y \otimes \hat{\boldsymbol{a}}_z$, and $\hat{\boldsymbol{n}} = \hat{\boldsymbol{n}}_y \otimes \hat{\boldsymbol{n}}_z$ are the reconstructed joint 2D steering vectors for the UE, BS, and RIS, respectively.

\subsection{Computational Complexity Analysis}

The computational complexity of the proposed methods is evaluated in terms of floating-point operations (Flops) and summarized in Table~\ref{tab:complexity_sbrt}. All estimators share the foundational costs of matched filtering ($KMQT$), path gain estimation ($QMN$), and 2D spatial frequency decoupling ($M+Q+N$). 

The Successive Kronecker Factorization (SKF) extracts the spatial signatures hierarchically via successive rank-one approximations. Conversely, the tensor-based estimators operate directly on the reshaped observation $\mathcal{E} \in \mathbb{C}^{Q \times M \times N}$. The iterative TOPE-ALS refines the factors over $I$ iterations, each costing $5QMN$ flops, whereas the deterministic TOPE-HOSVD computes the dominant singular vectors of the three principal mode-$n$ unfoldings without requiring iterations.

\begin{table}[htbp]
    \centering
    \caption{Computational Complexity of the Proposed Methods}
    \label{tab:complexity_sbrt}
    \resizebox{\columnwidth}{!}{%
    \begin{tabular}{@{}ll@{}}
        \toprule
        \textbf{Method} & \textbf{Total Cost (Flops)} \\ 
        \midrule
        SKF & $KMQT + 2QMN + MQ + M + Q + N$ \\
        TOPE-ALS & $KMQT + (5I + 1)QMN + M + Q + N$ \\
        TOPE-HOSVD & $KMQT + 4QMN + M + Q + N$ \\ 
        \bottomrule
    \end{tabular}%
    }
    \vspace{-0.5cm}
\end{table}

\begin{figure*}[t]
    \centering
    % Primeira figura ocupando 48% da largura total da página
    \begin{minipage}{0.48\textwidth}
        \centering
        \includegraphics[width=\linewidth]{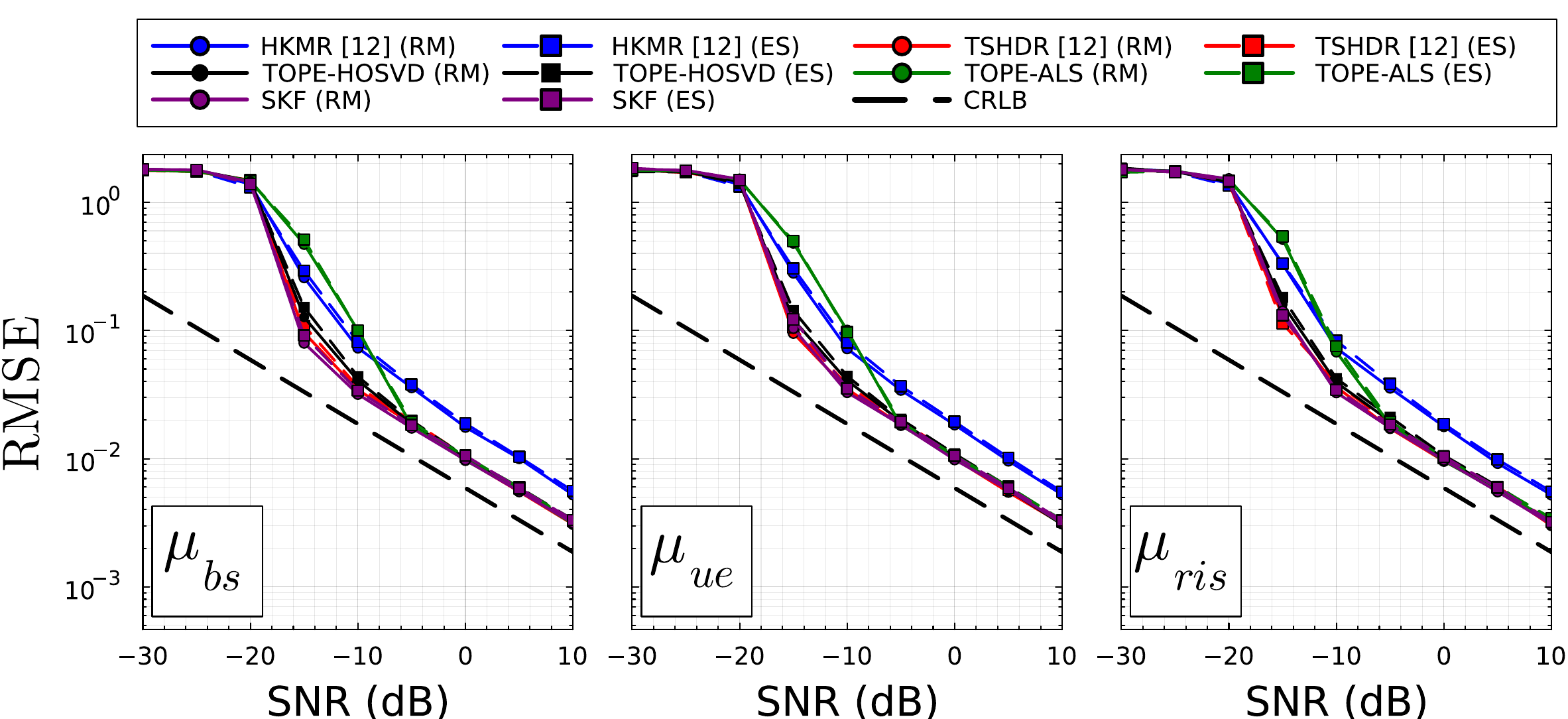}
        \caption{RMSE performance for the $\mu$ horizontal spatial frequencies at the BS, UE, and RIS.}
        \label{fig:rmse_mu_composta}
    \end{minipage}\hfill
    \vspace{-0.3cm}
    % Segunda figura ocupando os 48% restantes, com \hfill garantindo o espaçamento
    \begin{minipage}{0.48\textwidth}
        \centering
        \includegraphics[width=\linewidth]{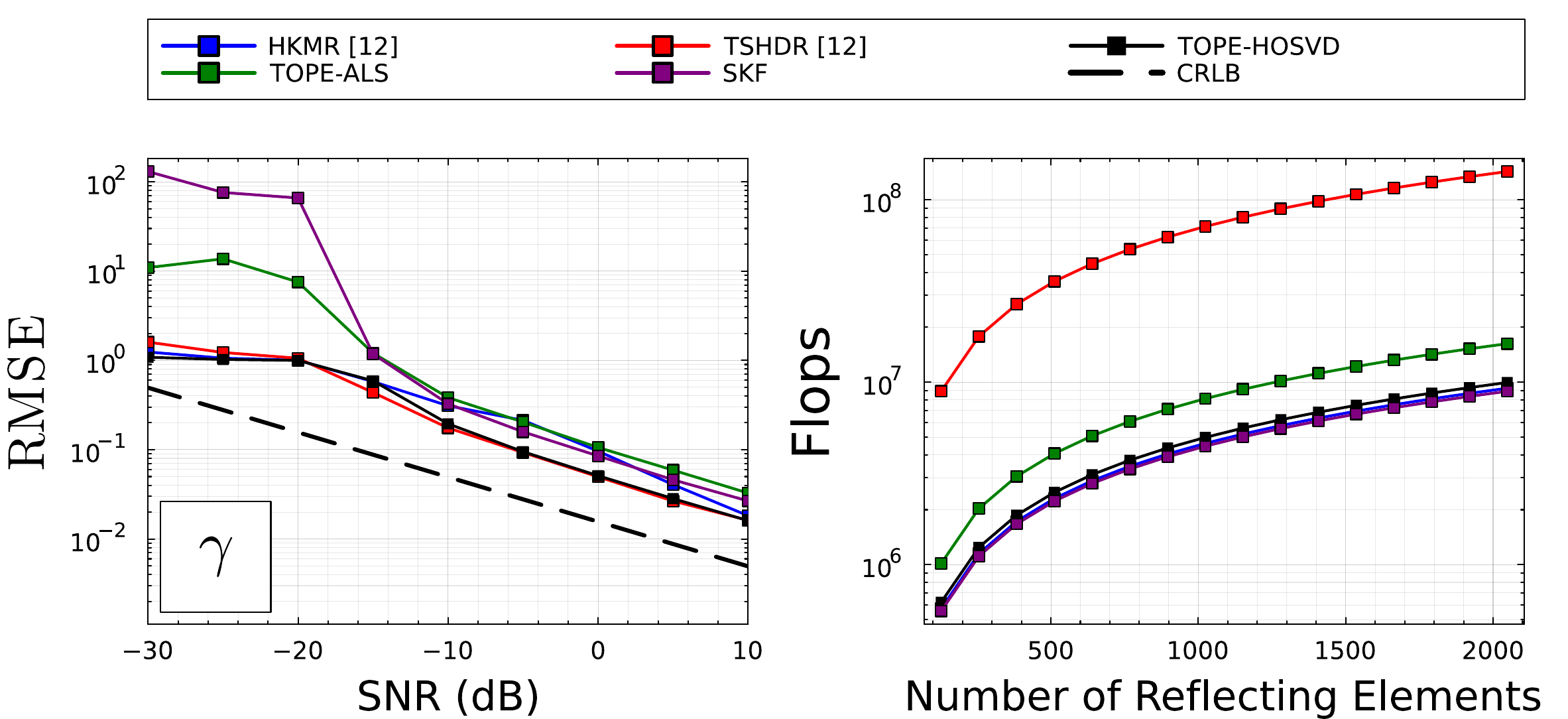}
        \caption{RMSE of the cascaded complex path gain $\gamma$ versus SNR, and overall computational complexity versus the number of RIS elements $N$.}
        \label{fig:ganho_complexidade}
    \end{minipage}
    \vspace{-0.3cm}
\end{figure*}
\section{Cramér-Rao Lower Bound}
This section derives the Cramér-Rao Lower Bound (CRLB) used as a theoretical benchmark for the proposed channel-parameter estimators. The unknown real-valued parameter vector is defined as $    \boldsymbol{\eta}=\left[\gamma_R,\gamma_I,\mu_{\text{bs}},\psi_{\text{bs}},\mu_{\text{ue}},\psi_{\text{ue}},\mu_y,\psi_z\right]^T$, where $\gamma=\gamma_R+j\gamma_I$ is the cascaded complex gain, $\{\mu_{\text{bs}},\psi_{\text{bs}}\}$ and $\{\mu_{\text{ue}},\psi_{\text{ue}}\}$ are the BS and UE spatial frequencies, and $\{\mu_y,\psi_z\}$ are the combined RIS spatial frequencies.

Let $\widehat{\boldsymbol{y}}=\operatorname{vec}(\widehat{\boldsymbol{Y}})\in\mathbb{C}^{QTK\times 1}$ denote the vector containing all received pilot samples over the $K$ RIS training blocks, and let $\boldsymbol{y}(\boldsymbol{\eta})=\operatorname{vec}(\boldsymbol{Y}(\boldsymbol{\eta}))$ be its noise-free counterpart. Under additive circularly symmetric complex Gaussian noise, the observation model is
\begin{equation}
    \widehat{\boldsymbol{y}}=\boldsymbol{y}(\boldsymbol{\eta})+\boldsymbol{v},
    \qquad
    \boldsymbol{v}\sim\mathcal{CN}(\boldsymbol{0},\boldsymbol{R}),
\end{equation}
where $\boldsymbol{R}=\sigma_n^2\boldsymbol{I}_{QTK}$ for spatially and temporally white noise. Hence, the conditional likelihood is
\begin{equation}
\begin{split}
    L(\widehat{\boldsymbol{y}};\boldsymbol{\eta})
    &=\frac{1}{\pi^{QTK}\det(\boldsymbol{R})}
    \exp\big[-\boldsymbol{r}^H(\boldsymbol{\eta})\boldsymbol{R}^{-1}\boldsymbol{r}(\boldsymbol{\eta})\big],
\end{split}
    \label{eq:likelihood}
\end{equation}
where $\boldsymbol{r}(\boldsymbol{\eta})=\widehat{\boldsymbol{y}}-\boldsymbol{y}(\boldsymbol{\eta})$ is the residual vector.
The log-likelihood is $\ell(\widehat{\boldsymbol{y}};\boldsymbol{\eta})=\ln L(\widehat{\boldsymbol{y}};\boldsymbol{\eta})$.

For the real-valued parameter vector $\boldsymbol{\eta}$ and white Gaussian noise with variance $\sigma_n^2$, the $(i,j)$-th entry of the Fisher Information Matrix (FIM) is
\begin{equation}
    [\boldsymbol{F}(\boldsymbol{\eta})]_{ij}
    =\frac{2}{\sigma_n^2}\operatorname{Re}\left\{
    \left(\frac{\partial\boldsymbol{y}(\boldsymbol{\eta})}{\partial\eta_i}\right)^H
    \left(\frac{\partial\boldsymbol{y}(\boldsymbol{\eta})}{\partial\eta_j}\right)
    \right\}.
    \label{eq:fim_entries}
\end{equation}
Therefore, for any unbiased estimate $\hat{\eta}_i$ of $\eta_i$, the CRLB establishes the following lower bounds on the variance and the root mean-square error (RMSE):
\begin{equation}
    \operatorname{var}(\hat{\eta}_i)\geq [\boldsymbol{F}^{-1}(\boldsymbol{\eta})]_{ii}, \quad \operatorname{RMSE}(\hat{\eta}_i)\geq \sqrt{[\boldsymbol{F}^{-1}(\boldsymbol{\eta})]_{ii}},
    \label{eq:crlb_bounds}
\end{equation}
where the latter establishes the corresponding lower bound on the root mean-square error (RMSE).

It remains to specify the deterministic vector $\boldsymbol{y}(\boldsymbol{\eta})$ and its derivatives. From \eqref{eq:rec_signal}, the noise-free received signal at the $k$-th RIS training block is
\begin{equation}
    \boldsymbol{Y}_k(\boldsymbol{\eta})=\boldsymbol{G}\operatorname{diag}(\boldsymbol{\omega}_k)\boldsymbol{H}\boldsymbol{S}\in\mathbb{C}^{Q\times T}.
\end{equation}

Using the property $\operatorname{vec}(\boldsymbol{ABC}) = (\boldsymbol{C}^T \otimes \boldsymbol{A})\operatorname{vec}(\boldsymbol{B})$, the vectorized noise-free signal for the $k$-th block is $\operatorname{vec}(\boldsymbol{Y}_k(\eta)) = (\boldsymbol{S}^T \otimes \boldsymbol{I}_Q)(\boldsymbol{H}^T \diamond \boldsymbol{G})\boldsymbol{\omega}_k$. Collecting all $K$ blocks into $\boldsymbol{Y}(\eta) \in \mathbb{C}^{QT \times K}$ and vectorizing across the training sequence, we obtain the complete observation vector:
\begin{equation}
    \boldsymbol{y}(\eta) = \operatorname{vec}(\boldsymbol{Y}(\eta)) = (\boldsymbol{\Omega}^T \otimes \boldsymbol{S}^T \otimes \boldsymbol{I}_Q)\operatorname{vec}(\boldsymbol{H}^T \diamond \boldsymbol{G}).
    \label{eq:y_eta_complete}
\end{equation}

Applying the vectorization operator to the rank-one factorization in \eqref{eq:kr_channel_factorization} yields $\operatorname{vec}(\boldsymbol{H}^T \diamond \boldsymbol{G}) = \gamma \, \boldsymbol{n} \otimes \boldsymbol{a} \otimes \boldsymbol{q}$. Thus, the complete noise-free observation vector can be written compactly as
\begin{equation}
    \boldsymbol{y}(\boldsymbol{\eta})
    = \gamma\boldsymbol{\Phi}\boldsymbol{c}(\boldsymbol{\eta}),
    \qquad
    \boldsymbol{\Phi}=\boldsymbol{\Omega}^T\otimes\boldsymbol{S}^T\otimes\boldsymbol{I}_Q,
    \label{eq:crlb}
\end{equation}
where $\boldsymbol{c}(\boldsymbol{\eta}) = \boldsymbol{n}(\mu_y,\psi_z)\otimes\boldsymbol{a}(\mu_{\text{bs}},\psi_{\text{bs}})\otimes\boldsymbol{q}(\mu_{\text{ue}},\psi_{\text{ue}})$.
    
    The combined RIS frequencies are $\mu_y=\mu_{\text{ris}_A}+\mu_{\text{ris}_D}$ and $\psi_z=\psi_{\text{ris}_A}+\psi_{\text{ris}_D}$. The derivatives required in \eqref{eq:fim_entries} follow directly from \eqref{eq:crlb}. In particular, $\frac{\partial\boldsymbol{y}}{\partial\gamma_R}=\boldsymbol{\Phi}\boldsymbol{c}(\boldsymbol{\eta}),\frac{\partial\boldsymbol{y}}{\partial\gamma_I}=j\boldsymbol{\Phi}\boldsymbol{c}(\boldsymbol{\eta})$, and, for any spatial-frequency parameter $\xi\in\{\mu_{\text{bs}},\psi_{\text{bs}},\mu_{\text{ue}},\psi_{\text{ue}},\mu_y,\psi_z\}$, $\frac{\partial\boldsymbol{y}}{\partial\xi} = \gamma\boldsymbol{\Phi}\frac{\partial\boldsymbol{c}(\boldsymbol{\eta})}{\partial\xi}$. For a one-dimensional steering vector $\boldsymbol{s}(\xi)=[1,e^{-j\xi},\ldots,e^{-j(L-1)\xi}]^T$, the derivative used in $\partial\boldsymbol{c}(\boldsymbol{\eta})/\partial\xi$ is
\begin{equation}
    \frac{\partial\boldsymbol{s}(\xi)}{\partial\xi}
    = -j\operatorname{diag}(0,1,\ldots,L-1)\boldsymbol{s}(\xi).
\end{equation}
These derivatives complete the construction of the FIM and, consequently, the CRLB for all entries of $\boldsymbol{\eta}$.
\section{Simulation Results}
\begin{table}[htbp]
    \centering
    \caption{Simulation Parameters}
    \label{tab:sim_parameters}
    \resizebox{\columnwidth}{!}{%
    \begin{tabular}{@{}lcc@{}}
        \toprule
        \textbf{Parameter} & \textbf{Symbol} & \textbf{Value} \\
        \midrule
        Array dimensions (BS, UE, RIS) & $M, Q, N$ & $16 \ (4 \times 4)$ \\
        Pilot sequence length & $T$ & $16$ \\
        Transmit power & $P_t$ & $1$ \\
        Azimuth angles & $\phi \in \{\text{bs}, \text{ris}_A, \text{ris}_D, \text{ue}\}$ & $\mathcal{U}(-60^{\circ}, +60^{\circ})$ \\
        Elevation angles & $\theta \in \{\text{bs}, \text{ris}_A, \text{ris}_D, \text{ue}\}$ & $\mathcal{U}(90^{\circ}, 130^{\circ})$ \\
        ALS tolerance \& Monte Carlo runs & $\epsilon$, runs & $10^{-12}$, $1000$ \\
        \bottomrule
    \end{tabular}%
    }
\end{table}

This section evaluates the estimation accuracy and computational complexity of the proposed estimators against the TSHDR~\cite{ref12} and HKMR~\cite{ref12} methods. The simulation setup follows the parameters listed in Table~\ref{tab:sim_parameters}.

The estimation accuracy is measured in terms of the root-mean-square error (RMSE) of the channel parameters. For a generic parameter $x$, the RMSE is defined as
\begin{equation}
    \operatorname{RMSE}(x)\!=\!\sqrt{\mathbb{E}\big[|x-\hat{x}|^2\big]}, x\!\in\!\{\gamma,\!\mu_{\text{bs}},\!\psi_{\text{bs}},\!\mu_{\text{ue}},\!\psi_{\text{ue}},\!\mu_y,\!\psi_z\}.
\end{equation}
In practice, the expectation is approximated by averaging over the Monte Carlo trials. The signal-to-noise ratio (SNR) is defined as $\mathrm{SNR} = P_t / \sigma_n^2$ where $\sigma_n^2$ is the noise variance.

Fig.~\ref{fig:rmse_mu_composta} evaluates the RMSEs of the horizontal spatial frequencies ($\mu$) for the BS, RIS, and UE; vertical frequencies ($\psi$) exhibit identical trends and are omitted. The curves closely track the CRLB, validating the estimators' efficiency and our derivations. Crucially, our estimators match the near-optimal TSHDR~\cite{ref12} benchmark, outperforming HKMR~\cite{ref12}. This confirms that early spatial-signature isolation preserves multidimensional array gains. While TOPE-ALS shows minor degradation at low SNRs due to slower convergence, it rapidly attains TSHDR-level precision. Furthermore, Root-MUSIC and ESPRIT yield nearly indistinguishable curves, indicating performance is governed by the recovered steering vectors rather than the chosen 1D estimator.

The complex-gain results in Fig.~\ref{fig:ganho_complexidade} corroborate the spatial frequency analysis, with the proposed estimators overlapping the TSHDR~\cite{ref12} and HKMR~\cite{ref12} curves. However, the proposed framework's true advantage lies in breaking the accuracy--complexity compromise (Fig.~\ref{fig:ganho_complexidade}, right). The HKMR~\cite{ref12} method achieves a low complexity by applying a Kronecker factorization directly on the received pilots, a premature decoupling that discards joint spatial structures and degrades accuracy. Conversely, the TSHDR~\cite{ref12} method preserves this gain via matched filtering but applies the decoupling on the massive global channel matrix, resulting in a prohibitive $\mathcal{O}(M^2 Q^2 N)$ complexity.

By restricting Kronecker factorization to the extracted lower-dimensional signature vectors, our estimators reduce the dominant complexity to $\mathcal{O}(QMN)$. Consequently, SKF matches HKMR's footprint, while TOPE-HOSVD undercuts it. Even accounting for iterations, TOPE-ALS remains vastly more efficient than TSHDR.

Finally, regarding practical limitations, our estimators are structurally tailored to the rank-1 algebraic formulation of an ideal single-path LoS channel. In more challenging propagation environments, the cascaded channel may present sparse multi-path components, which could be accommodated by extending our framework to higher-rank tensor models, akin to the geometric approach in [16]. Furthermore, addressing the impact of non-ideal operational conditions, such as hardware impairments and imperfect phase control, as recently highlighted in unstructured RIS estimation frameworks like [17], represents a critical and highly promising direction for future research.

\section{Conclusions}
This paper proposed three efficient channel estimators (SKF, TOPE-ALS, and TOPE-HOSVD) for RIS-assisted MIMO systems employing URAs under a dominant LoS model. By isolating multidimensional spatial signatures prior to frequency decoupling, our framework breaks the accuracy-complexity tradeoff. Analysis confirm these estimators attain the near-optimal accuracy of TSHDR~\cite{ref12} with a minimal computational footprint comparable to HKMR~\cite{ref12}. Furthermore, our exact CRLB derivation proves they operate close to fundamental theoretical limits. While tailored to rank-1 LoS environments, extending this factorization to higher-rank tensor models for multi-path resolution remains a promising future direction.

% \section*{Aknowledgements} %##################################
% The Coordination of the SBrT~2026 thanks to the Coordination of the previous symposiums promoted by the Brazilian Telecommunication Society, for making available this example.

% \appendix %###################################################
% Insert informations about the apendix here


\scriptsize
\begin{thebibliography}{99}
\sloppy
\bibitem{ref1} Q. Wu and R. Zhang, ``Towards smart and reconfigurable environment: Intelligent reflecting surface aided wireless network,'' \textit{IEEE Communications Magazine}, vol. 58, no. 1, pp. 106--112, Jan. 2020.

\bibitem{ref2} M. A. ElMossallamy, H. Zhang, L. Song, K. G. Seddik, G. Y. Li, and Z. Han, ``Reconfigurable intelligent surfaces for wireless communications: Principles, challenges, and opportunities,'' \textit{IEEE Transactions on Cognitive Communications and Networking}, vol. 6, no. 3, pp. 990--1002, Sept. 2020.

\bibitem{ref3} Y. Yu, J. Wang, X. Zhou, C. Wang, and Z. Bai, ``Review on channel estimation for reconfigurable intelligent surface assisted wireless communication system,'' \textit{Mathematics}, vol. 11, no. 14, Art. no. 3235, 2023.

\bibitem{ref4} A. L. Swindlehurst, G. Zhou, R. Liu, C. Pan, and M. Li, ``Channel estimation with reconfigurable intelligent surfaces---A general framework,'' \textit{Proceedings of the IEEE}, vol. 110, no. 9, pp. 1464--1488, Sept. 2022.

\bibitem{ref5} E. Basar, M. Di Renzo, J. de Rosny, M. Debbah, M.-S. Alouini, and R. Zhang, ``Wireless communications through reconfigurable intelligent surfaces,'' \textit{IEEE Access}, vol. 7, pp. 116753--116773, 2019.

\bibitem{ref6} C. You, B. Zheng, and R. Zhang, ``Channel estimation and passive beamforming for intelligent reflecting surface: Discrete phase shift and progressive refinement,'' \textit{IEEE Journal on Selected Areas in Communications}, vol. 38, no. 11, pp. 2604--2620, Nov. 2020.

\bibitem{ref7} G. T. de Ara\'ujo, A. L. F. de Almeida, and R. Boyer, ``Channel estimation for intelligent reflecting surface assisted MIMO systems: A tensor modeling approach,'' arXiv:2008.04766, 2020.

\bibitem{ref8} H. Xiao, H. Deng, A. Guo, Y. Qian, C. Peng, and Y. Zhang, ``Accelerated PARAFAC-based channel estimation for reconfigurable intelligent surface-assisted MISO systems,'' \textit{Sensors}, vol. 22, no. 19, Art. no. 7463, 2022.

\bibitem{ref9} T. G. Kolda and B. W. Bader, ``Tensor decompositions and applications,'' \textit{SIAM Review}, vol. 51, no. 3, pp. 455--500, 2009.

\bibitem{ref10} N. D. Sidiropoulos, L. De Lathauwer, X. Fu, K. Huang, E. E. Papalexakis, and C. Faloutsos, ``Tensor decomposition for signal processing and machine learning,'' \textit{IEEE Transactions on Signal Processing}, vol. 65, no. 13, pp. 3551--3582, July 2017.

\bibitem{ref11} K. B. A. Ben\'icio, A. L. F. de Almeida, B. Sokal, F.-E. Asim, B. Makki, and G. Fodor, ``Tensor-based modeling/estimation of static channels in IRS-assisted MIMO systems,'' in \textit{Proc. XLI Brazilian Symposium on Telecommunications and Signal Processing (SBrT)}, 2023.

\bibitem{ref12} 
Fazal-E-Asim, A. L. F. de Almeida, B. Sokal, B. Makki e G. Fodor, ``Two-Dimensional Channel Parameter Estimation for IRS-Assisted Networks,'' \textit{IEEE Transactions on Communications}, vol. 73, no. 8, pp. 6337-6350, Ago. 2025. doi: 10.1109/TCOMM.2024.3522047.

\bibitem{ref13}
A. Vesa, ``Direction of Arrival Estimation using MUSIC and RootMUSIC Algorithm,'' in \textit{Proc. 18th Telecommunications Forum (TELFOR)}, Belgrade, Serbia, Nov. 2010.

\bibitem{ref14}
R. K and P. K. N, ``Performance Evaluation \& Analysis of Direction of Arrival Estimation Algorithms using ULA,'' in \textit{2018 International Conference on Electrical, Electronics, Communication, Computer, and Optimization Techniques (ICEECCOT)}, Mysuru, India, 2018, pp. 1467-1473. doi: 10.1109/ICEECCOT43722.2018.9001455.


\bibitem{ref15}
F.-E. Asim, B. Sokal, A. L. F. de Almeida, B. Makki, and G. Fodor, ``Structured channel estimation for RIS-assisted THz communications,'' \textit{IEEE Transactions on Vehicular Technology}, vol. 74, no. 3, pp. 5175--5180, Mar. 2025, doi: 10.1109/TVT.2024.3492998.

\bibitem{ref16}
F.-E. Asim, A. L. F. de Almeida, B. Sokal, B. Makki, and G. Fodor, ``Deconstructing the composite channel for beyond diagonal RIS: Channel estimation and beamforming design,'' \textit{arXiv preprint arXiv:2606.01564}, 2026.

\bibitem{ref17}
D. V. C. de Oliveira, D. C. Alcantara, F.-E. Asim, A. L. F. de Almeida, and G. Fodor, ``Simultaneous direct and indirect channel estimation for RIS-assisted MIMO communications,'' in \textit{XLIII Simpósio Brasileiro de Telecomunicações e Processamento de Sinais (SBrT)}, 2025. doi: 10.14209/sbrt.2025.1571157164.

\end{thebibliography}
\end{document}